\documentstyle{article}
\font\tenbf=cmbx10
\font\elevenbf=cmbx10 scaled\magstep 1
\font\elevenrm=cmr10 scaled\magstep 1
\font\elevenit=cmti10 scaled\magstep 1
\font\ninerm=cmr9
\renewenvironment{thebibliography}[1]
 { \elevenrm
   \begin{list}{\arabic{enumi}.}
    {\usecounter{enumi} \setlength{\parsep}{0pt}
     \setlength{\itemsep}{3pt} \settowidth{\labelwidth}{#1.}
     \sloppy
    }}{\end{list}}

\def\beee{\begin{equation}}
\def\eeee{\end{equation}}
\def\dggg{^{\dagger}}

\begin{document}
\begin{center}{{\tenbf (PARA)BOSONS, (PARA)FERMIONS, QUONS AND\\
\vglue 3 pt
OTHER BEASTS IN THE MENAGERIE OF PARTICLE STATISTICS}
\footnote{\ninerm\baselineskip=11pt {Talk presented by
O.W. Greenberg at the International Conference on Fundamental Aspects
of Quantum Theory to celebrate the 60th birthday of Yakir Aharonov.}}
\vglue 1.0cm
{\tenrm O.W. GREENBERG\\}
\baselineskip=13pt
{\tenit Center for Theoretical Physics,
Department of Physics, University of
Maryland\\}
\baselineskip=12pt
{\tenit College Park, MD 20742-4111, USA,\\}
\vglue 0.3cm
{\tenrm D.M. GREENBERGER\\}
{\tenit Department of Physics, City College of the City University of
New York\\ New York, NY 10031, USA\\}
\vglue 0.2cm
{\tenrm and\\}
\vglue 0.2cm
{\tenrm T.V. GREENBERGEST\\}
{\tenit Department of Physics, Southern Methodist University,
Dallas, TX 75275, USA\\}
\vglue 0.3cm
{\tenrm University of Maryland Preprint No. 93-203}
\vglue 0.8cm
{\tenrm ABSTRACT}}
\end{center}
\vglue 0.3cm
{\rightskip=3pc
 \leftskip=3pc
 \tenrm\baselineskip=12pt
 \noindent
After some general comments about statistics and the TCP theorem, I discuss
experimental searches for violations of the exclusion principle and
theories which allow for such violations.
\vglue 0.6cm}
{\elevenbf\noindent 1. Introduction}
\vglue 0.2cm
\baselineskip=14pt
\elevenrm
It is a great pleasure to speak at this symposium honoring Yakir Aharonov.
Because of the broad range of Yakir's interests, I have been able to see people
who work in different areas than mine whom I don't usually see at conferences
and to meet for the first time people whose names and work I know, but whom I
had never had the opportunity to meet.  Yakir is especially concerned with
fundamental issues which have lasting interest, such as particle statistics.
In
the first part of my talk I will say some things about statistics and related
issues which may not be generally known, and in the second part I will focus on
how well we know that particles obey the statistics we think they obey and on
theories which allow violations of statistics.

By way of introduction, I mention two relations involving spin which are
on quite different footings.  The relation between spin and isospin, that
integer-spin particles have integer isospin and odd-half-integer-spin particles
have odd-half-integer isospin, was suggested on the basis of few examples: the
proton and neutron, which are in the odd-half-integer category and the three
pions, which are in the integer category.  Further, there was no fundamental
basis for such a relation.  When strange particles were discovered, this
relation was found to be violated by the kaons, which have zero spin and
isospin
one-half, and by the lambda and sigma hyperons, which have spin one-half and
integer isospin.  Since there was no theory supporting this relation, it was
easy to discard it.  By contrast, the relation between spin and statistics
first
stated by Pauli$^1$ in 1936, that integer-spin particles obey Bose
statistics and
odd-half-integer-spin particles obey Fermi statistics was supported by many
examples and, at least for free fields, was proved by Pauli from the basic
requirement of local commutativity of observables.  This relation has survived
and is one of the most general results of quantum field theory.
\vglue 0.6cm
{\elevenbf\noindent 2. General Comments about Statistics and Related Issues}
\vglue 0.2cm
{\elevenit\noindent 2.1 Additivity of the Energy of Widely Separated
Subsystems}
\vglue 0.1cm
\baselineskip=14pt
\elevenrm
The zeroth condition I discuss is the requirement that the energy of widely
separated subsystems be additive.  This requires that all terms in the
Hamiltonian be ``effective Bose operators'' in that sense that
\elevenit
\beee
[{\cal H}({\bf x}), \phi({\bf y})]_-\rightarrow 0, |{\bf x}-{\bf y}|\rightarrow
\infty.
\eeee
\elevenrm
For example,  ${\elevenit \cal H}$ can't have a term such as
$\phi(x){\elevenit \psi(x)}$, where
${\elevenit \phi}$ is
Bose and $\psi$ is Fermi, because then the contributions to the
energy of widely separated subsystems would alternate in sign.  Such terms are
also prohibited by rotational symmetry.
\vglue 0.2cm
{\elevenit \noindent 2.2 Statistics of Bound States is Determined by
Statistics of Constituents}
\vglue 0.1cm
The well-known rule that a bound state of any number of Bosons and an even
number of Fermions is a Boson, while a bound state with an odd number of
Fermions is a Fermion, was first stated by Wigner,$^2$ who published in
Hungarian
and suffered the consequence of using a relatively inaccessible language.
Later Ehrenfest and Oppenheimer$^3$ independently
published this result in English.
\vglue 0.6cm
{\elevenit\noindent 2.3 Spin-statistics Theorem}
\vglue 0.1cm
\baselineskip=14pt
\elevenrm
I distinguish between two theorems.  The {\elevenit physical}
spin-statistics theorem
is the theorem of Pauli mentioned above, local commutativity of observables
requires that, given the choice between Bose and Fermi statistics, integer-spin
particles must obey Bose statistics and odd-half-integer-spin particles must
obey Fermi statistics.  The phrase, {\elevenit given the choice between}, is
necessary, because the analogous connection holds between parabose or parafermi
statistics and spin.  The theorem which I prefer to call the
spin--type-of-locality theorem, due to Burgoyne,$^4$
states that fields which commute at
spacelike separation must have integer spin and fields that anticommute at
spacelike separation must have odd-half-integer spin.  Both the assumptions and
the conclusions of the two theorems differ.  The Pauli theorem explicitly
assumes a choice between different types of {\elevenit particle}
statistics and concludes
that if the wrong choice is made, then observables fail to commute at spacelike
separation.  For example, if one chooses Bose statistics for spin-one-half
particles, i.e., uses Bose commutation relations for the annihilation and
creation operators of the spin-one-half particles, then the commutator of the
observables for the free theory will contain the $S^{(1)}(x-y)$ singular
function,
which does not vanish for spacelike $x-y$, rather than the $S(x-y)$ singular
function which does.  The theory (at least for the free case) still exists.
The Burgoyne
theorem makes no statement about particle statistics; rather it assumes a
choice
between {\elevenit field} commutation rules.  If the wrong choice is
made, then the fields are identically zero, so the theory does not even exist.
This latter theorem has a very general proof in the context of axiomatic field
theory; however it says nothing about {\elevenit particle} statistics.
\vglue 0.2cm
{\elevenit \noindent 2.4 Weakness of the TCP Theorem}
\vglue 0.1cm
In contrast to the spin-statistics theorem, which requires locality of
observables, the TCP theorem holds regardless of locality, and is a much weaker
theorem.  Indeed, it is difficult to make a theory which violates TCP.  This is
clearly illustrated by Jost's example.$^5$  Jost shows that a free
neutral scalar field whose annihilation and creation operators are quantized
with anticommutation relations (and whose particles thus obey Fermi statistics)
still obeys the
{\elevenit normal} TCP theorem.  Cluster decomposition properties also
hold regardless of the choice of commutation relations.
\vglue 0.4cm
{\elevenbf\noindent 3. Search for Small Violations of Fermi and Bose
Statistics}
\vglue 0.4cm
Now I come to the second part of my talk and discuss how to detect violations
of
Fermi or Bose statistics if they occur.
Atomic spectroscopy is the first place to search for violations of the
exclusion principle since that is where
Pauli discovered it.  One looks for funny lines which
do not correspond to lines in the normal theory of atomic spectra.  There are
such lines, for example in the solar spectrum; however they probably can be
accounted for in terms of highly ionized atoms in an environment of high
pressure, high density and large magnetic fields.  Laboratory spectra are well
accounted for by theory and can bound the violation of the exclusion principle
for electrons by something like $10^{-6}$ to $10^{-8}$.  A useful
quantitative measure of the violation, ${\cal V}$, is that ${\cal V}$ is the
coefficient of
the anomalous component of the two-particle density matrix; for fermions,
the two-electron density matrix, $\rho_2$, is
\beee
\rho_2=(1-{\cal V}) \rho_a+{\cal V} \rho_s,
\eeee
where $\rho_{a(s)}$ is the antisymmetric (symmetric) two-fermion density
matrix.
Thoma and Nolte,$^6$ in
a contribution to a poster session here, discuss bounds on the violation of the
exclusion principle for nucleons based on the absence of the nucleus $^5Li$.
Bounds also follow from the absence of $^5He$.  Mohapatra and I surveyed a
variety of searches for violations of particle statistics in $^7$.

I will discuss an insightful experiment by Maurice and Trudy
Goldhaber$^8$ which was designed to answer the question, ``Are the
electrons emitted in nuclear
$\beta$-decay quantum mechanically identical to the
electrons in atoms?''  We know that the $\beta$-decay electrons have the same
spin, charge and mass as electrons in atoms; however the Goldhabers realized
that if the $\beta$-decay electrons were not
quantum mechanically identical to those
in atoms, then the $\beta$-decay
electrons would not see the K shell of a heavy
atom as filled and would fall into the K shell and emit an x-ray.  The
Goldhabers looked for such x-rays by letting $\beta$-decay electrons from a
natural source fall on a block of lead.  No such x-rays were found.  The
Goldhabers were able to confirm that electrons from the two sources are indeed
quantum mechanically identical.  At the same time, they found that any
violation of the exclusion principle for electrons must be less than $5\%$.

Ramberg and Snow$^{9}$ developed this experiment into one which yields a
high-precision bound on violations of the exclusion principle.  Their idea was
to replace the natural $\beta$ source, which provides relatively few electrons,
by an electric current, in which case Avogadro's number is on our side.  The
possible violation of the exclusion principle is that a given collection of
electrons can, with different probabilities, be in different permutation
symmetry states.  The probability to be in the ``normal'' totally antisymmetric
state would presumably be close to one,
the next largest probability would occur for the
state with its Young tableau having one row with two boxes, etc.  The idea of
the experiment is that each collection of electrons has a possibility of being
in an ``abnormal'' permutation state.  If the density matrix for a conduction
electron together with the electrons in an atom has a projection onto such an
``abnormal'' state, then the conduction electron will not see the K shell of
that atom as filled. Then a transition into the K shell with x-ray emission
is allowed.  Each conduction electron which comes sufficiently close to a given
atom has an independent chance to make such an x-ray-emitting transition, and
thus the probability of seeing such an x-ray is proportional to the number of
conduction electrons which traverse the sample and the number of atoms which
the
electrons visit, as well as the probability that a collection of electrons can
be in the anomalous state.  Ramberg and Snow chose to run 30 amperes
through a thin copper strip for about a month.  They estimated the energy of
the
x-rays which would be emitted due to the transition to the K shell.  No
excess of x-rays above background was found in this energy region.  Ramberg
and Snow set the limit
\beee
{\cal V} \leq 1.7 \times 10^{-26}.
\eeee
This is high precision, indeed!
\vglue 0.6cm
{\elevenbf \noindent 4. Theories of Violation of Statistics}
\vglue 0.2cm
{\elevenit \noindent 4.1 Gentile's Intermediate Statistics}
\vglue 0.1cm
The first attempt to go beyond Bose and Fermi statistics seems to have been
made by G. Gentile$^{10}$  who suggested an
``intermediate statistics'' in which at
most $n$ identical particles could occupy a given quantum state.  In
intermediate
statistics, Fermi statistics is recovered for $n=1$ and Bose statistics
is recovered for $n\rightarrow \infty$; thus intermediate statistics
interpolates between Fermi and Bose statistics.  However, Gentile's
statistics is not a proper quantum statistics, because the condition of having
at most $n$ particles in a given quantum state is not invariant under change
of basis.  For example, for intermediate statistics with $n=2$, the state
$|\psi \rangle=|k,k,k \rangle$ does not exist; however, the state $|\chi
\rangle=
\sum_{l_1,l_2,l_3}U_{k,l_1}U_{k,l_2}U_{k,l_3}|l_1,l_2,l_3 \rangle$, obtained
from $|\psi \rangle$ by the unitary change of single-particle basis,
$|k \rangle ^{\prime}=\sum_{l}U_{k,l}|l \rangle$ does exist.

By contrast, parafermi statistics of order $n$ is
invariant under change of basis.$^{11}$  Parafermi statistics of order
$n$ not only
allows at most $n$ identical particles in the same state, but also allows
at most $n$ identical particles in a symmetric state.  In the example just
described, neither $|\psi \rangle$ nor $|\chi \rangle$ exist for parafermi
statistics of order two.
\vglue 0.2cm
{\elevenit \noindent 4.2 Green's Parastatistics}
\vglue 0.1cm
H.S. Green$^{12}$ proposed the first proper quantum statistical
generalization of Bose and Fermi statistics.  Green noticed that the commutator
of the number operator with the annihilation and creation operators is the same
for both bosons and fermions
\beee
[n_k, a\dggg_l]_-=\delta_{kl}a\dggg_l.
\eeee
The number operator can be written
\beee
n_k=(1/2)[a\dggg_k, a_k]_{\pm}+ {\rm const},
\eeee
where the anticommutator (commutator) is for the Bose (Fermi) case.  If these
expressions are inserted in the number operator-creation operator commutation
relation, the resulting relation is {\elevenit trilinear}
in the annihilation and creation operators.  Polarizing the number operator to
get the transition operator $n_{kl}$ which annihilates a free particle in state
$k$ and creates one in state $l$ leads to Green's trilinear commutation
relation
for his parabose and parafermi statistics,
\beee
[[a\dggg_k, a_l]_{\pm}, a\dggg_m]_-=2\delta_{lm}a\dggg_k
\eeee
Since these rules are trilinear, the usual vacuum condition,
\beee
a_k|0\rangle=0,
\eeee
does not suffice to allow calculation of matrix elements of the $a$'s and
$a\dggg$'s; a condition on one-particle states must be added,
\beee
a_k a\dggg_l|0\rangle=\delta_{kl}|0\rangle.
\eeee

Green found an infinite set of solutions of his commutation rules, one for each
integer, by giving an ansatz which he expressed in terms of Bose and Fermi
operators.  Let
\beee
a_k\dggg=\sum_{p=1}^n b_k^{(\alpha) \dagger},~~a_k=\sum_{p=1}^n b_k^{(\alpha)},
\eeee
and let the $b_k^{(\alpha)}$ and $b_k^{(\beta) \dagger}$
be Bose (Fermi) operators
for $\alpha=\beta$ but anticommute (commute) for $\alpha \neq \beta$ for the
``parabose'' (``parafermi'') cases.  This ansatz clearly satisfies Green's
relation.  The integer $p$ is the order of the parastatistics.  The physical
interpretation of $p$ is that, for parabosons, $p$ is the maximum number of
particles that can occupy an antisymmetric state, while for parafermions, $p$
is the maximum number of particles that can occupy a symmetric state (in
particular, the maximum number which can occupy the same state).  The case
$p=1$
corresponds to the usual Bose or Fermi statistics.
Later, Messiah
and I$^{11}$ proved that Green's ansatz gives all Fock-like solutions of
Green's commutation rules.  Local observables have a form analogous to the
usual
ones; for example, the local current for a spin-1/2 theory is
$j_{\mu}=(1/2)[\bar{\psi}(x), \psi(x)]_-$.  From Green's ansatz, it is clear
that the squares of all norms of states are positive, since sums of Bose or
Fermi operators give positive norms.  Thus parastatistics gives a set of
orthodox theories.  Parastatistics is one of the
possibilities found by Doplicher, Haag and Roberts$^{13}$ in a general study of
particle statistics using algebraic field theory methods.  A good review of
this
work is in Haag's recent book$^{14}$.

This is all well and good; however, the violations of statistics provided by
parastatistics are gross.  Parafermi statistics of order 2 has up to 2
particles in each quantum state.  High-precision experiments are not necessary
to rule this out for all particles we think are fermions.
\vglue 0.2cm
{\elevenit \noindent 4.3 The Ignatiev-Kuzmin Model and ``Parons''}
\vglue 0.1cm
Interest in possible small violations of the exclusion principle was revived by
a paper of Ignatiev and Kuzmin$^{15}$ in 1987.  They constructed a model of
one oscillator with three possible states: a vacuum state, a one-particle
state and, with small probability, a two-particle state.  They gave trilinear
commutation relations for their
oscillator.  Mohapatra and I showed that the Ignatiev-Kuzmin oscillator could
be
represented by a modified form of the order-two Green ansatz.  We suspected
that
a field theory generalization of this model having an infinite number of
oscillators
would not have local observables and set
about trying to prove this.  To our surprize, we found that we could construct
local observables and gave trilinear relations which guarantee the locality
of the current.$^{16}$  We also checked the positivity of the norms with
states of three or less particles.  At this stage, we were carried away with
enthusiasm, named these particles ``parons'' since their algebra is a
deformation of the parastatistics algebra, and thought we had found a local
theory with small violation of the exclusion principle.
We did not know that Govorkov$^{17}$ had shown in
generality that any deformation of the Green commutation relations necessarily
has states with negative squared norms in the Fock-like representation.
For our model, the first such negative-probability state occurs for
four particles in the representation of ${\cal S}_4$ with three boxes in the
first row and one in the second.  We were able to understand Govorkov's result
qualitatively as follows:$^{18}$
Since parastatistics of order $p$ is related by a
Klein transformation to a model with exact $SO(p)$ or $SU(p)$ internal
symmetry,
a deformation of parastatistics which interpolates between Fermi and parafermi
statistics of order two would be equivalent to interpolating between the
trivial
group whose only element is the identity and a theory with
$SO(p)$ or $SU(p)$ internal symmetry.  This is impossible, since there is no
such interpolating group.
\vglue 0.2cm
{\elevenit \noindent 4.4 Apparent Violations of Statistics Due to
Compositeness}
\vglue 0.1cm
Before getting to ``quons,'' the final type of statistics I will discuss, I
want
to interpolate some comments about apparent violations of statistics due to
compositeness.  Consider two $^3 He$ nuclei, each of which is a fermion.  If
these two nuclei are brought in close proximity, the exclusion principle will
force each of them into excited states, plausibly with small amplitudes for the
excited states.  Let the creation operator for the nucleus at location $A$ be
\beee
b_A\dggg=\sqrt{1-\lambda_A^2}b_0\dggg+\lambda_A b_1\dggg+\cdots, |\lambda_A|<<1
\eeee
and the creation operator for the nucleus at location $B$ be
\beee
b_B\dggg=\sqrt{1-\lambda_B^2}b_0\dggg+\lambda_B b_1\dggg+\cdots,
|\lambda_B|<<1.
\eeee
Since these nuclei are fermions, the creation operators obey fermi statistics,
\beee
[b_i\dggg, b_j\dggg]_+=0
\eeee
Then,
\beee
b_A\dggg b_B\dggg|0\rangle
=[\sqrt{1-\lambda_A^2}\lambda_B-\lambda_A\sqrt{1-\lambda_B^2}]
b_0\dggg b_1\dggg |0\rangle,
\eeee
\beee
\|b_A\dggg b_B\dggg|0\rangle \|^2 \approx (\lambda_A-\lambda_B)^2<<1,
\eeee
so, with small probability,
the two could even occupy the same location, because each could be excited
into higher states with different amplitudes.  This is not an intrinsic
violation of the exclusion principle, but rather only an apparent violation due
to compositeness.
\vglue 0.2cm
{\elevenit \noindent 4.5 ``Quons''}
\vglue 0.1cm
Now I come to my last topic, ``quons.''$^{19}$  The quon algebra is
\beee
a_k a_l \dggg-q a_l \dggg a_k=\delta_{kl}.
\eeee
For the Fock-like representation which I consider, the vacuum condition
\beee
a_k |0\rangle=0
\eeee
is imposed.

These two conditions determine all vacuum matrix element of polynomials in the
creation and annihilation operators.  In the case of free quons, all
non-vanishing vacuum matrix elements must have the same number of annihilators
and creators.  For such a matrix element with all annihilators to the left and
creators to the right, the matrix element is a sum of products of
``contractions'' of the form $\langle 0|a a\dggg |0 \rangle$ just as in the
case
of bosons and fermions.  The only difference is that the terms are multiplied
by
integer powers of $q$.  The power can be given as a graphical rule:  Put
$\circ$'s
for each annihilator and $\times$'s for each creator in the order in which
they occur in the matrix element on the x-axis.
Draw lines above the x-axis connecting the pairs which are
contracted.  The minimum number of times these lines cross
is the power of $q$ for that term in the matrix element.

The physical significance of $q$ for small violations of Fermi statistics is
that $q=2 {\cal V} -1$, where the parameter ${\cal V}$ appears in Eq.(2).
For small violations of Bose statistics, the two-particle density matrix is
\beee
\rho_2=(1-{\cal V}) \rho_s+{\cal V} \rho_a,
\eeee
where $\rho_{s(a)}$ is the symmetric (antisymmetric) two-boson density matrix.
Then $q=1- 2{\cal V}$.

For $q$ in the open interval $(-1, 1)$ all
representations of the symmetric group occur.  As $q \rightarrow 1$, the
symmetric representations are more heavily weighted and at $q=1$
only the totally
symmetric representation remains; correspondingly, as $q \rightarrow -1$, the
antisymmetric representations are more heavily weighted and at $q=-1$ only the
totally antisymmetric representation remains.  Thus for a general
$n$-quon state, there
are $n!$ linearly independent states for $-1<q<1$, but there is only one
state for $q= \pm 1$.
I emphasize something that many people find very strange: {\elevenit there is
no
commutation relation between two creation or between two annihilation
operators,} except for $q= \pm 1$, which, of course, correspond to Bose and
Fermi statistics.  Indeed, the fact that the general $n$-particle state with
different quantum numbers for all the particles has $n!$ linearly independent
states proves that there is no such commutation relation between any number
of creation
(or annihilation) operators. An even stronger statement holds:  There is no
two-sided ideal containing a term with only creation operators.
Note that here quons differ from the ``quantum plane'' in which
\beee
xy=qyx
\eeee
holds.

Quons are an operator realization of ``infinite statistics'' which were found
as
a possible statistics by Doplicher, Haag and Roberts$^{13}$ in their general
classification of particle statistics.  The simplest case, $q=0$,$^{20}$,
suggested to me by Hegstrom,
was discussed earlier in the context of operator algebras by Cuntz.$^{21}$
It seems
likely that the Fock-like representations of quons for $|q|<1$ are homotopic to
each other and, in particular, to the $q=0$ case, which is particularly simple.
Thus it is convenient, as I will now do,
to illustrate qualitative properties of quons for this simple case.  All
bilinear observables can be constructed from the number operator,
$n_k \equiv n_{kk}$, or the
transition operator, $n_{kl}$, which obey
\beee
[n_k, a\dggg_l]_-=\delta_{kl}a\dggg_l,
{}~~[n_{kl}, a\dggg_m]_-=\delta_{lm}a\dggg_k
\eeee
Although the formulas for $n_k$ and $n_{kl}$ in the general case$^{22}$ are
complicated, the corresponding formulas for $q=0$ are simple.$^{20}$
Once Eq.(18) holds, the Hamiltonian and other observables can be
constructed in the usual way; for example,
\beee
H=\sum_k \epsilon_k n_k,~~ {\rm etc.}  \label{8}
\eeee
The obvious thing is to try
\beee
n_k=a\dggg_k a_k.   \label{9}
\eeee
Then
\beee
[n_k,a\dggg_l]_-=a\dggg_ka_ka\dggg_l-a\dggg_la\dggg_ka_k.  \label{10}
\eeee
The first term in Eq.(\ref{10}) is $\delta_{kl}a\dggg_k$ as desired; however
the second term is extra and must be canceled.  This can be done by adding the
term $\sum_ta\dggg_ta\dggg_ka_ka_t$ to the term in Eq.(\ref{9}).  This cancels
the extra term, but adds a new extra term, which must be canceled by another
term.  This procedure yields an infinite series for the number operator
and for the transition operator,
\beee
n_{kl}=a\dggg_ka_l+\sum_ta\dggg_ta\dggg_ka_la_t+\sum_{t_1,t_2}a\dggg_{t_2}
a\dggg_{t_1}a\dggg_ka_la_{t_1}a_{t_2}+ \dots   \label{11}
\eeee
As in the Bose case, this infinite series for the transition or number
operator defines an unbounded operator whose domain includes states made by
polynomials in the creation operators acting on the vacuum.
(As far as I know, this is the first case in which the number operator,
Hamiltonian, etc. for a free field are of infinite degree.  Presumably this is
due to the fact that quons are a deformation of an algebra and are related to
quantum groups.)
For nonrelativistic theories, the x-space form of the transition operator
is$^{23}$
\begin{eqnarray}
\rho_1({\bf x};{\bf y})=\psi\dggg({\bf x})\psi({\bf y})
+\int d^3z\psi\dggg
({\bf z})\psi\dggg({\bf x})\psi({\bf y})\psi({\bf z})  \nonumber \\
+\int d^3z_1d^3z_2\psi({\bf
z_2})\psi\dggg({\bf z_1})\psi\dggg({\bf x})\psi({\bf y})\psi({\bf z_1})
\psi({\bf z_2})+ \cdots,  \label{121}
\end{eqnarray}
which obeys the nonrelativistic locality requirement
\beee
[\rho_1({\bf x};{\bf y}),\psi\dggg({\bf w})]_-=\delta({\bf y}-{\bf
w})\psi\dggg({\bf x}),~~ {\rm and}~~
\rho({\bf x};{\bf y})|0\rangle=0.  \label{12}
\eeee
The apparent nonlocality of this formula associated with the space integrals
has
no physical significance.  To support this last statement, consider
\beee
[Qj_{\mu}(x),Qj_{\nu}(y)]_-=0,~~x \sim y,   \label{13}
\eeee
where $Q=\int d^3x j^0(x)$.  Equation (\ref{13}) seems to have nonlocality
because of the space integral in the $Q$ factors; however, if
\beee
[j_{\mu}(x),j_{\nu}(y)]_-=0,~~x \sim y, \label{14}
\eeee
then Eq.({\ref{13}) holds, despite the apparent nonlocality.  What is relevant
is the commutation relation, not the representation in terms of a space
integral.  (The apparent nonlocality of quantum electrodynamics in the Coulomb
gauge is another such example.)

In a similar way,
\beee
[\rho_2({\bf x}, {\bf y};{\bf y}^{\prime}, {\bf x}^{\prime}),
\psi\dggg({\bf z})]_-=\delta({\bf x}^{\prime}-{\bf z})
\psi\dggg({\bf x})\rho_1({\bf y},{\bf y}^{\prime})+
\delta({\bf y}^{\prime}-{\bf z})
\psi\dggg({\bf y})\rho_1({\bf x},{\bf x}^{\prime}).   \label{15}
\eeee
Then the Hamiltonian of a nonrelativistic theory with two-body interactions has
the form
\beee
H=(2m)^{-1} \int d^3x \nabla _x \cdot \nabla_{x^{\prime}}
\rho_1({\bf x}, {\bf x}^{\prime})|_{{\bf x}={\bf x}^{\prime}} +
\frac{1}{2} \int d^3x d^3y V(|{\bf x}-{\bf y}|)
\rho_2({\bf x},{\bf y};{\bf y}, {\bf x}).              \label{16}
\eeee
\begin{eqnarray}
[H,\psi\dggg({\bf z}_1) \dots \psi\dggg({\bf z}_n)]_-=
[-(2m)^{-1} \sum_{j=1}^n \nabla^2_{{\bf z}_i}+
 \sum _{i<j}
V(|{\bf z}_i-{\bf z}_j|)] \psi\dggg({\bf z}_1) \dots \psi\dggg({\bf z}_n)
\nonumber \\
+\sum_{j=1}^n \int d^3x V(|{\bf x}-{\bf z}_j|)\psi\dggg({\bf z}_1)
\cdots \psi\dggg({\bf z}_n)\rho_1({\bf x}, {\bf x}^{\prime}).   \label{17}
\end{eqnarray}
Since the last term on the right-hand-side of Eq.(\ref{17}) vanishes when the
equation is applied to the vacuum, this equation shows that the usual
Schr\"odinger equation holds for the $n$-particle system.  Thus the usual
quantum mechanics is valid, with the sole exception that any permutation
symmetry is allowed for the many-particle system.  This construction justifies
calculating the energy levels of (anomalous) atoms with electrons in states
which violate the exclusion principle using the normal Hamiltonian,
but allowing anomalous
permutation symmetry for the electrons.$^{24}$

I have not yet addressed the question of positivity of the squares of norms
which caused grief in the paron model.  Several authors have given proofs of
positivity.$^{25-28}$  The proof of Zagier provides an
explicit formula for the determinant of the $n! \times n!$ matrix
of scalar products among the states of $n$ particles in different quantum
states.  Since this determinant is one for $q=0$, the norms will be positive
unless the determinant has zeros on the real axis.  Zagier's formula
\beee
det\; M_{n}(q) =\Pi_{k=1}^{n-1}(1-q^{k(k+1)})^{(n-k)n!/k(k+1)},
\eeee
has zeros only on the unit circle, so the desired positivity follows.
Although quons satisfy the requirements of nonrelativistic locality, the quon
field does not obey the relativistic requirement, namely, spacelike
commutativity of observables.  Since quons interpolate smoothly between
fermions, which must have odd half-integer spin, and bosons, which must have
integer spin, the spin-statistics theorem, which can be proved, at least for
free fields, from locality would be violated if locality were to hold for quon
fields.  It is amusing that, nonetheless, the free quon field obeys the TCP
theorem and Wick's theorem holds for quon fields.$^{19}$

It is well known that external fermionic sources must be multiplied by a
Grassmann number in order to be a valid term in a Hamiltonian.  This is
necessary, because additivity of the energy of widely separated systems
requires
that all terms in the Hamiltonian must be effective Bose operators.  I
constructed the quon analog of Grassmann numbers$^{29}$ in order to allow
external quon sources.  Because this issue was overlooked, the bound on
violations of Bose statistics for photons claimed
in$^{30}$ is invalid.
\vglue 0.2cm
{\elevenit \noindent 4.6 Speicher's ansatz}
\vglue 0.1cm
Speicher$^{27}$ has given an ansatz for the Fock-like representation of quons
analogous to Green's ansatz for parastatistics.  Speicher represents the quon
annihilation operator as
\beee
a_k={\rm lim}_{N \rightarrow \infty}N^{-1/2}\sum_{\alpha=1}^N b_k^{(\alpha)},
\eeee
where the $b_k^{(\alpha)}$ are Bose oscillators for each $\alpha$, but with
relative commutation relations given by
\beee
b_k^{(\alpha)} b_l^{(\beta) \dagger}=s^{(\alpha, \beta)}b_l^{(\beta) \dagger}
b_k^{(\alpha)}, \alpha \neq \beta,~ {\rm where}~~ s^{(\alpha, \beta)}= \pm 1.
\eeee
This limit is taken as the limit, $N \rightarrow \infty$,
in the vacuum expectation state of the
Fock space representation of the
$b_k^{(\alpha)}$.
In this respect, Speicher's ansatz differs from Green's, which is an operator
identity.  Further, to get the Fock-like representation of the quon algebra,
Speicher chooses a probabilistic condition for the signs
$s^{(\alpha, \beta)}$,
\beee
{\rm prob}(s^{(\alpha, \beta)}=1)=(1+q)/2,
\eeee
\beee
{\rm prob}(s^{(\alpha, \beta)}=-1)=(1-q)/2.
\eeee
Rabi Mohapatra and I tried to get a specific ansatz for the
$s^{(\alpha, \beta)}$ without success.  I was concerned about that, but
Jonathan Rosenberg, one of
my mathematical colleagues at Maryland, pointed out that some things which are
easy to prove on a probabilistic basis are difficult to prove otherwise.  For
example, it is easy to prove that with probability one any number is
transcendental, but difficult to prove that $\pi$ is transcendental.  I close
my
discussion of Speicher's ansatz with two comments.  First, one must assume the
probability distribution is uncorrelated, that is
\beee
(1/N^2)\sum_{\alpha, \beta} s^{(\alpha, \beta)}=[(1/2)(1+q)(1)+(1/2)(1-q)(-1)]
=q,
\eeee
\beee
(1/N^3)\sum_{\alpha, \beta, \gamma} s^{(\alpha, \beta)}s^{(\beta), \gamma)}
=q^2
\eeee
\beee
(1/N^3)\sum_{\alpha, \beta, \gamma}s^{(\alpha, \beta)}s^{(\beta, \gamma)}
s^{(\gamma, \alpha)}=q^3,
\eeee
etc.  Secondly, one might think
that, since Eq.(32) implies the analogous relation
for two annihilators or two creators in the Fock-like representation,
Speicher's ansatz would imply $a_k a_l-q a_l a_k=0$, which we know cannot hold.
This problem would arise if the ansatz were an operator identity, but does not
arise for the limit in the Fock vacuum.  Since a sum of Bose operators acting
on
a Fock vacuum always gives a positive-definite norm, the positivity property is
obvious with Speicher's construction.

Speicher's ansatz leads to the conjecture that there is an
infinite-valued hidden degree of freedom underlying $q$-deformations analogous
to the hidden degree of freedom underlying parastatistics.
\vglue 0.6cm
{\elevenbf \noindent 5. Acknowledgements}
\vglue 0.4cm
I thank the Graduate Research Board, University of Maryland, College Park,
for a Semester Research Award and the University for a Sabbatical Leave during
which this work was carried out.  This work was supported in part by
the National Science Foundation.
\vglue 0.5cm
{\elevenbf\noindent 6. References \hfil}
\vglue 0.4cm


\begin{thebibliography}{99}
\bibitem{fivel} D.I. Fivel, {\elevenit Phys. Rev. Lett.} {\elevenbf 65} (1990)
3361; erratum, {\elevenit ibid} {\elevenbf 69} (1992) 2020. I have not
checked that the erratum repairs the error in the proof of positivity in the
article.
\end{thebibliography}
\end{document}